\documentclass[prl,twocolumn,showpacs]{revtex4}
\usepackage[T1]{fontenc}
\usepackage{amsmath}
\usepackage{graphics}
\usepackage{amssymb}
\usepackage{pslatex}

\makeatletter

\providecommand{\LyX}{L\kern-.1667em\lower.25em\hbox{Y}\kern-.125emX\@}

\makeatother
\begin{document}

\title{Low-energy Properties of Aperiodic Quantum Spin Chains}

\author{Andr\'e P. Vieira}

\affiliation{Instituto de F\'{\i}sica da Universidade de S\~{a}o Paulo, Caixa Postal 66318,
05315-970, S\~{a}o Paulo, Brazil}

\email{apvieira@if.usp.br}

\date{\today}


\begin{abstract}
We investigate the low-energy properties of antiferromagnetic quantum
\emph{XXZ} spin chains with couplings following two-letter aperiodic
sequences, by an adaptation of the Ma-Dasgupta-Hu renormalization-group
method. For a given aperiodic sequence, we argue that in the easy-plane
anisotropy regime, intermediate between the \emph{XX} and Heisenberg
limits, the general scaling form of the thermodynamic properties is
essentially given by the exactly known \emph{XX} behavior, providing
a classification of the effects of aperiodicity on \emph{XXZ} chains.
As representative illustrations, we present analytical and numerical
results for the low-temperature thermodynamics and the ground-state
correlations for couplings following the Fibonacci quasiperiodic structure
and a binary Rudin-Shapiro sequence, whose geometrical fluctuations
are similar to those induced by randomness.
\end{abstract}
\maketitle
At low temperatures, the interplay between lack of translational invariance
and quantum fluctuations in low-dimensional strongly correlated electron
systems may induce novel phases with peculiar behavior. In particular,
randomness in quantum spin chains may lead, for instance, to Griffiths
phases \cite{fisher92}, large-spin formation \cite{westerberg95},
and random-singlet phases \cite{fisher94}.
On the other hand, studies on the influence of deterministic but aperiodic
elements on similar systems (see e.g. \cite{luck86,luck93a,hermisson00,hida99,hida01,vidal99}),
inspired by the experimental discovery of quasicrystals,
have revealed strong effects on dynamical and thermodynamic properties,
but far less is known concerning the precise nature of the underlying
ground-state phases.

Prototypical models for those studies are spin-\( \frac {1}{2} \)
antiferromagnetic (AFM) \emph{XXZ} chains described by the Hamiltonian\begin{equation}
H=\sum _{i}J_{i}\left( S^{x}_{i}S^{x}_{i+1}+S^{y}_{i}S^{y}_{i+1}+\Delta S^{z}_{i}S^{z}_{i+1}\right) ,
\end{equation}
where all \( J_{i}>0 \) and the \( S_{i} \) are spin operators.
Random-bond versions of these systems have been much studied by a
real-space renormalization-group (RG) method introduced \cite{ma79}
by Ma, Dasgupta and Hu (MDH) for the Heisenberg chain (\( \Delta =1 \))
and more recently extended by Fisher \cite{fisher92,fisher94}, who
gave evidence that the method becomes asymptotically exact at low
energies. The idea is to decimate the spin pairs coupled by the strongest
bonds (those with the largest gaps between the local ground state and the
first excited multiplet), forming singlets and inducing weak effective
couplings between neighboring spins, thereby reducing the energy scale.
For \emph{XXZ} chains in the regime \( -\frac {1}{2} < \Delta \leq 1 \),
the method predicts the ground state to be a random-singlet phase,
consisting of arbitrarily distant spins forming rare, strongly correlated
singlet pairs \cite{fisher94}. Here we employ the MDH method to investigate
the low-energy properties of aperiodic chains.

Two-letter aperiodic sequences (AS's) can be generated by an inflation
rule such as \( a\rightarrow ab \), \( b\rightarrow a \), which
produces the Fibonacci sequence \( abaababa\ldots  \). Associating
with each \( a \) a coupling \( J_{a} \) and with each \( b \)
a coupling \( J_{b} \) we can build an aperiodic quantum spin chain.
In the \emph{XX} limit (\( \Delta =0 \)),
the low-temperature thermodynamic behavior can be qualitatively
determined for any AS by an
exact RG method \cite{hermisson00}. The effects of aperiodicity
depend on topological properties of the AS.
If the fraction of letters \(a\) (or \(b\)) at odd positions is different
from that at even positions (i.e., if there is average dimerization), then a finite gap opens between the global ground state and the first excited states, and the chain becomes noncritical.
Otherwise, the scaling of the lowest gaps can be classified according to the wandering exponent
\( \omega  \) measuring the geometric fluctuations \( g \) related
to nonoverlapping pairs of letters \cite{hermisson00}, which vary
with the system size \( N \) as \( g\sim N^{\omega } \). If \( \omega <0 \),
aperiodicity has no effect on the long-distance, low-temperature properties,
and the system behaves as in the uniform case, with a finite susceptibility
at \( T=0 \). If \( \omega =0 \), as in the Fibonacci sequence,
aperiodicity is marginal and may lead to nonuniversal power-law scaling
behavior of thermodynamic properties. If \( \omega >0 \), aperiodicity
is relevant in the RG sense, affecting the \( T=0 \) critical behavior
and leading to exponential scaling of the lowest gaps (\( \Lambda  \))
at long distances \( r \), of the form \( \Lambda \sim \exp \left( -r^{\omega }\right)  \).
In particular, for sequences with \( \omega =\frac {1}{2} \),
geometric fluctuations mimic those induced by randomness, and the
scaling behavior is similar to the one characterizing the random-singlet
phase \cite{fisher94}. No analogous results exist for general \emph{XXZ}
chains, although bosonization and density-matrix RG (DMRG) calculations on the Heisenberg
chain indicate that Fibonacci couplings
should be relevant
\cite{vidal99,hida99,hida01}. We argue below that, for a given AS, low-temperature properties of all chains in the regime \( 0\leq \Delta \leq 1 \)
should follow essentially the \emph{XX} scaling form. Moreover, we also obtain information on ground-state correlation functions.

In order to apply the MDH method to aperiodic chains, we must remember
that now there are many spin blocks with the same gap at a given
energy scale. Also, those blocks may consist of more than two
spins, in which case effective spins would form upon renormalization.
The strategy is to sweep through the lattice until
all blocks with the same gap have been renormalized, leading to new
effective couplings (and possibly spins). Then we search for the next
largest gap, which again corresponds to many blocks. When all possible
original blocks have been considered, there remains some unrenormalized
spins, possibly along with effective ones, defining new blocks which
form a second generation of the lattice. The process is then iterated,
leading to the renormalization of the spatial distribution of
effective blocks along the generations; for small enough coupling
ratios (which are indeed required for the
MDH method to work), \emph{this distribution will be the same for}
\emph{all} \( 0\leq \Delta \leq 1 \). Because of the self-similarity
inherent to AS's generated by inflation rules, the effective couplings
take a finite number of values, and it is
natural that the block distribution reaches a periodic attractor
(usually a fixed point) after a few lattice sweeps.
By studying recursion relations for the effective couplings, we can obtain analytical results.
As the RG proceeds, the coupling ratio usually gets smaller, suggesting that
the method becomes asymptotically exact.
This picture holds for marginal (\(\omega =0\)) and relevant (\(\omega >0\)) aperiodicity; for irrelevant AS's, such as the Thue-Morse sequence
(\(a\rightarrow ab\), \(b\rightarrow ba\)), the coupling ratio approaches unity as the RG proceeds, and the method eventually breaks down.
As representative examples, we consider the marginal
case of Fibonacci couplings and the relevant case of a binary Rudin-Shapiro
sequence. Full details of the calculations, as well as application
to other AS's, will be reported elsewhere \cite{vieira04}.

The blocks to be renormalized consist of \( n \) spins connected
by equal bonds \( J_{0} \), and coupled to the rest of the chain
by weaker bonds \( J_{l} \) and \( J_{r} \). The ground state (GS)
for blocks with an even number of spins is a singlet, and at low energies we can eliminate the whole block,
along with \( J_{l} \) and \( J_{r} \), leaving an effective AFM
bond \( J^{\prime } \) coupling the two spins closer to the block
and given by second-order perturbation theory as \( J^{\prime }=\gamma _{n}J_{l}J_{r}/J_{0} \),
with \( \Delta  \)-dependent coefficients \( \gamma _{n} \). A block
with an odd number of spins has a doublet as its GS; at
low energies, it can be replaced by an effective spin connected to
its nearest neighbors by AFM effective bonds \( J_{l,r}^{\prime }=\gamma _{n}J_{l,r} \)
whose values are calculated by first-order perturbation theory.
In general, the anisotropy parameters are
also renormalized and become site-dependent; for \( n \) even, the
effective anisotropy is \( \Delta ^{\prime }=\delta _{n}(\Delta _{0})\Delta _{l}\Delta _{r} \),
while for \( n \) odd \( \Delta ^{\prime }_{l,r}=\delta _{n}(\Delta _{0})\Delta _{l,r} \),
with \( \left| \delta _{n}(\Delta )\right| <1 \) for \( 0\leq \Delta <1 \)
and \( \delta _{n}(1)=1 \). Thus, for \( 0<\Delta <1 \) the \( \Delta _{i} \)
flow to the \emph{XX} fixed point (all \( \Delta _{i}=0 \)), ultimately
reproducing the corresponding scaling behavior, while for the Heisenberg
chain all \( \Delta _{i} \) remain equal to unity. So, we focus here
on the Heisenberg and \emph{XX} limits, and postpone examples for
intermediate cases to a future publication \cite{vieira04}.

\begin{figure}
{\centering \resizebox*{1\columnwidth}{!}{\includegraphics{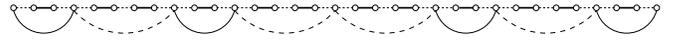}} \par}
\caption{\label{figfib}Left end of the Fibonacci \emph{XXZ} chain. Dashed
(solid) lines represent \protect\( J_{a}\protect \) (\protect\( J_{b}\protect \))
bonds. An effective coupling \protect\( J^{\prime }_{b}\protect \)
is induced between spins separated by only one singlet pair, while
\protect\( J^{\prime }_{a}\protect \) connects spins separated by
two singlet pairs. Apart from a few bonds close to the chain ends,
the effective couplings also form a Fibonacci sequence.}
\end{figure}

First we apply the method to chains with Fibonacci couplings. This
is the simplest example of the quasiperiodic precious-mean sequences
with marginal fluctuations \cite{hermisson00}. A few bonds closer
to the left end of the original chain, along with induced effective
couplings, are shown in Fig. \ref{figfib} for \( J_{a}<J_{b} \) \cite{note4}.
Only singlets are formed by the RG process, producing two different
effective couplings, \[
J^{\prime }_{a}=\gamma ^{2}_{2}J^{3}_{a}/J^{2}_{b}\quad \text {and}\quad J^{\prime }_{b}=\gamma _{2}J^{2}_{a}/J_{b}.\]
 The bare coupling ratio is \( \rho =J_{a}/J_{b} \), its renormalized
value being \( \rho ^{\prime }=\gamma _{2}\rho  \). In each generation
\( j \), all decimated blocks have the same size \( r_{j} \) and
gap \( \Lambda _{j} \) (proportional to the effective \( J_{b} \)
bonds). The recursion relations for \( \rho  \) and \( \Lambda  \)
are given by\[
\rho _{j+1}=\gamma _{2}\rho _{j}\quad \text {and}\quad \Lambda _{j+1}=\gamma _{2}\rho ^{2}_{j}\Lambda _{j}.\]
%whose solution yields\[
%\Lambda _{j+1}=\gamma ^{j^{2}}_{2}\rho ^{2j}J_{b}.\]
The distance between spins forming a singlet in the \( j \)th generation
defines a characteristic length \( r_{j} \), corresponding to the
Fibonacci numbers \( r_{j}=1 \), \( 3 \), \( 13 \), \( 55 \),
\( \ldots  \); for \( j\gg 1 \) the ratio \( r_{j+1}/r_{j} \) approaches
\( \phi ^{3} \), where \( \phi =(1+\sqrt{5})/2 \) is the golden
mean. So we have
\( r_{j}\sim r_{0}\phi^{3j}\),
%\( j=\ln (r_{j}/r_{0})/3\ln \phi  \),
where \( r_{0} \)
is a constant, and by solving the recursion relations we obtain the dynamic scaling behavior,
\begin{equation}
\label{gapheisfib}
\Lambda _{j}\sim r_{j}^{-\zeta }e^{-\mu \ln ^{2}(r_{j}/r_{0})},
\end{equation}
with \( \zeta =-\frac {2}{3}\ln \rho /\ln \phi  \) and \( \mu =-\ln \gamma _{2}/9\ln ^{2}\phi  \).
For the Heisenberg chain \( \gamma _{2}=\frac {1}{2} \), and
Eq. (\ref{gapheisfib}) describes a weakly exponential scaling, but
not of the form \( \Lambda \sim \exp \left( -r^{\omega }\right)  \)
found for the \emph{XX} chain with relevant aperiodicity (\( \omega >0 \))
and used to fit the DMRG data for the Fibonacci Heisenberg chain \cite{hida99,hida01}.
For the \emph{XX} chain \( \gamma _{2}=1 \), so that \( \mu =0 \)
and we can identify \( \zeta  \) with a dynamical critical exponent
\( z \), whose value depends on the coupling ratio, leading to nonuniversal
scaling behavior, characteristic of strictly marginal operators \cite{note7}.
This nonuniversality should hold in the anisotropy
regime \( 0<\Delta <1 \) with a ``bare'' value of \( \rho  \) defined
at a crossover scale.
Note that we can view the Heisenberg scaling form (\( \mu \neq 0 \))
as a marginally relevant \( (\omega \rightarrow 0^{+} \)) case.

\begin{figure}
{\centering \resizebox*{0.85\columnwidth}{!}{\includegraphics{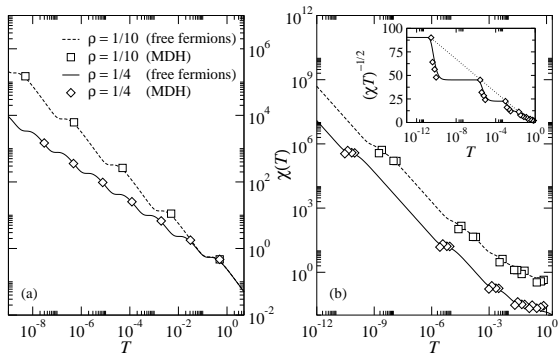}} \par}
\caption{\label{figsusc}Behavior of \protect\( \chi (T)\protect \) for aperiodic
\emph{XX} chains, obtained from the MDH method (symbols) and from
numerical diagonalization on chains with \protect\( 10^{4}\protect \)-\protect\( 10^{5}\protect \)
sites (curves), for two values of \protect\( \rho =J_{a}/J_{b}\protect \).
(a) Fibonacci couplings. The slope of the curves depends on \protect\( \rho \protect \),
reflecting the marginal character of the aperiodicity.
(b) Rudin-Shapiro
couplings. The inset plots the inverse square root of
\protect\( T\chi (T)\protect \) versus \protect\( T\protect \) (in
log scale) with \protect\( \rho =\frac {1}{4}\protect \), showing
that for points corresponding to the smallest gaps in each generation
the random-singlet phase result \protect\( \chi (T)\sim 1/T\ln ^{2}T\protect \)
is reproduced (dotted line).}
\end{figure}
The susceptibility \( \chi (T) \) can be estimated \cite{fisher94}
by assuming that, at energy scale \( \Lambda _{j}\sim T \), only
unrenormalized spins are magnetically active (and essentially free),
singlet pairs being effectively frozen. Thus, if \( n_{j}\sim r^{-1}_{j} \)
is the number of surviving spins in the \( j \)th generation, \( \chi (T\sim \Lambda _{j})\sim n_{j+1}/\Lambda _{j}. \)
As shown in Fig. \ref{figsusc}(a), \( \chi (T) \) estimated for
the Fibonacci \emph{XX} chain from the MDH method agrees very well
with results from free-fermion \cite{lieb61} numerical diagonalization (ND) of finite chains, even for \( \rho =\frac {1}{4} \).

As all singlets formed at the \( j \)th generation have size \( r_{j} \)
and the block distribution is fixed, the average GS
correlation between spins separated by a distance \( r_{j} \) is
\begin{equation}
\label{corrfib}
C^{\alpha \alpha }(r_{j})\equiv \overline{\left\langle S^{\alpha }_{i}S^{\alpha }_{i+r_{j}}\right\rangle }=\tfrac {1}{2}\left| c_{0}\right| \left( n_{j}-n_{j+1}\right) =\sigma \left| c_{0}\right| r^{-1}_{j},
\end{equation}
where \( \sigma  \) is a constant, \( \alpha =x,y,z \), and \( c_{0} \)
is the GS correlation between the two spins in a singlet, given by
\( c_{0}=-\frac {1}{4} \) for the Heisenberg chain and for both
\( \alpha =x \) and \( \alpha =z \) in the \emph{XX} chain. We point
out that these should be the dominant correlations, and spins separated
by distances other than \( r_{j} \) are predicted to be only weakly
correlated. This can be checked for the \emph{XX} chain by calculating
GS correlations via the free-fermion method. Results for chains with
\( \rho \lesssim \frac {1}{4} \) (see \cite{vieira04}) reveal
very good agreement with Eq. (\ref{corrfib}). As correlations in
the uniform \emph{XX} chain \cite{lieb61} decay as \( C^{xx}(r)\sim r^{-1/2} \)
and \( C^{zz}(r)\sim r^{-2} \), dominant \( xx \) (\( zz \)) correlations
in the Fibonacci chain are weaker (stronger) than in the uniform chain.

Relevant aperiodicity is characterized by strong geometric fluctuations,
and is usually induced by sequences with blocks having two or more neighboring
strong bonds. Furthermore,
after the first lattice sweep, more than two values of effective couplings
may be produced. However, they all derive from the original pair of
couplings \( J_{a} \) and \( J_{b} \), so that, in the presence of a periodic attractor, it is generally possible to write recursion
relations for an effective coupling ratio and gap having the forms
\cite{vieira04}
\begin{equation}
\label{rrrel}
\rho _{j+1}=c\rho ^{k}_{j}\quad \text {and}\quad \Lambda _{j+1}=\lambda \rho ^{\ell }_{j}\Lambda _{j},
\end{equation}
where \( c \) and \( \lambda  \) are \( \Delta  \)-dependent nonuniversal
constants, and \( \ell  \) (a rational number) and \( k \) (an integer)
relate to the number of singlets involved in determining the effective
couplings. We assume \( k\geq 2 \), \( k=1 \) corresponding to marginal
behavior, as in the Fibonacci case \cite{note3}. If a characteristic length scale
takes the form \( r_{j}=r_{0}\tau ^{j} \), with a rescaling factor
\( \tau  \), solving the recursion relations leads to a dynamic scaling
described by\begin{equation}
\label{dsfrel}
\Lambda _{j}\sim r^{-\zeta }_{j}\exp \left( -\mu r_{j}^{\omega }\right) \sim \exp \left( -\mu r_{j}^{\omega }\right) ,
\end{equation}
with \( \zeta  \) and \( \mu  \) nonuniversal constants and \( \omega =\ln k/\ln \tau  \).
Note that \( \omega  \) has \emph{precisely the same form} as the
exact result \cite{hermisson00} obtained for the \emph{XX} chain
with aperiodic couplings not inducing average dimerization. Moreover,
\( \omega  \) depends only on the topology of the sequence and on
its self-similar properties, but not on the anisotropy; so, for a
given AS, the scaling form in Eq. (\ref{dsfrel})
should be valid for any \emph{XXZ} chain in the regime \( 0\leq \Delta \leq 1 \),
with the same exponent \( \omega  \).

\begin{figure}
{\centering \resizebox*{0.8\columnwidth}{!}{\includegraphics{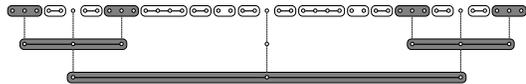}} \par}
\caption{\label{figrs}Left end of the first 3 generations of the Rudin-Shapiro
chains. Thick lines indicate strong bonds. Shaded blocks contribute
effective spins when renormalized; white blocks form singlets.}
\end{figure}
As an example, we consider couplings following the binary Rudin-Shapiro
(RS) sequence, whose inflation rule for letter pairs is \( aa\rightarrow aaab \),
\( ab\rightarrow aaba \), \( ba\rightarrow bbab \) and \( bb\rightarrow bbba \).
This generates blocks having between \( 2 \) and \( 5 \) spins.
Figure \ref{figrs} shows a few bonds closer to the left end of the
chain for the first three generations of the lattice. Blocks with
more than \( 3 \) spins are eliminated in the first sweep and do
not appear in later generations. Both two- and three-spin blocks are
present in the fixed-point block distribution (already reached at
the second generation), and
a hierarchy of effective spins is produced,
as depicted in Fig. \ref{figrs}. At the \( j \)th generation,
three-spin blocks have size \( r_{j}=2\times 4^{j-1} \) (so that \( \tau =4 \)),
while two-spin blocks have size \( r_{j}/2 \). The first lattice
sweep generates effective couplings \( \tilde{J}_{i} \) having 8
different values, but three of them are enough to write recursion
relations, \[
\tilde{J}^{\prime }_{a}=\gamma ^{2}_{2}\gamma _{3}\tilde{J}_{c}\tilde{J}^{2}_{a}/\tilde{J}^{2}_{b},\: \: \: \:
\tilde{J}^{\prime }_{b}=\gamma _{3}\tilde{J}_{c}, \]
\[\; \: \text {and}\; \: \tilde{J}^{\prime }_{c}=\gamma _{2}\gamma _{3}\tilde{J}_{a}\tilde{J}_{c}/\tilde{J}_{b}.\]
The gap in a given generation is proportional to the effective \( \tilde{J}_{b} \);
defining \( \rho =\tilde{J}_{a}/\tilde{J}_{b} \) and eliminating
\( \tilde{J}_{c} \) gives \[
\rho _{j+1}=\gamma ^{2}_{2}\rho ^{2}_{j}\quad \text {and}\quad \Lambda _{j+1}=\gamma ^{3}_{3}\rho ^{1/2}_{j}\Lambda _{j},\]
which correspond to the forms in Eq. (\ref{rrrel}) with \( k=2 \)
and \( \ell =1/2 \). So we obtain, for the whole regime \( 0\leq \Delta \leq 1 \),
the dynamical scaling form in Eq. (\ref{dsfrel}) with an exponent
\( \omega =\frac {1}{2} \), as predicted for the \emph{XX} chain,
reproducing the result for the random-singlet phase. In Fig. \ref{figsusc}(b)
we plot \( \chi (T) \) for the \emph{XX} chain calculated from both
ND and the MDH method, again obtaining very
good agreement.

\begin{figure}
{\centering \resizebox*{0.98\columnwidth}{!}{\includegraphics{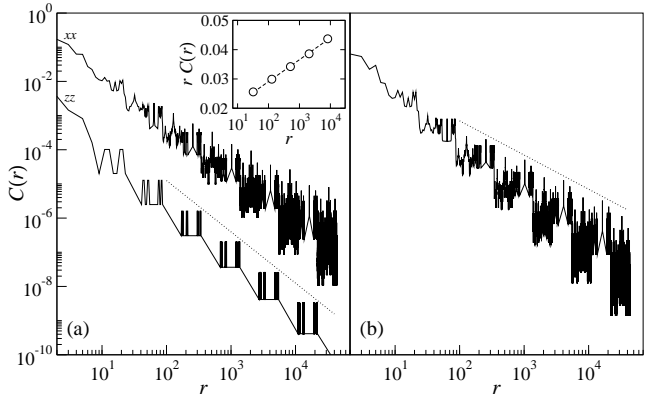}} \par}
\caption{\label{figrscorr}GS correlations for chains with RS couplings,
obtained from extrapolation of numerical MDH results for chains with
\protect\( 2^{16}\protect \) to \protect\( 2^{20}\protect \) sites.
(a) \protect\( C^{xx}(r)\protect \) (upper solid curve) and \protect\( C^{zz}(r)\protect \)
(lower solid curve) for the \emph{XX} chain. The dotted curve is proportional
to \protect\( r^{-3/2}\protect \). (Curves offset for clarity.) Inset:
dominant \protect\( C^{xx}\protect \) correlations, fitted by
a law of the form \protect\( rC^{xx}(r)=y_{0}+y_{1}\ln r\protect \)
(dashed curve). (b) \protect\( C(r)\protect \) for the Heisenberg
chain (solid curve). The dotted curve is proportional to \protect\( 1/r\protect \).}
\end{figure}
For chains with RS couplings, effective-spin formation from three-spin
blocks determines the dominant GS correlations. These blocks stem
from both original three-spin (and some five-spin) blocks and unrenormalized
spins.
An effective spin represents all spins in the original block via Clebsch-Gordan
coefficients, allowing us to calculate correlations between
any two spins whose effective spins end up in the same block at some
stage of the RG process.
Because of the hierarchical structure seen in Fig. \ref{figrs},
for each block renormalized at the \( j \)th generation the correlation
between its end spins connects a number of order \( 2^{j} \) original
spin pairs separated by the same distance \( r_{j} \) (the size of
the block), yielding a contribution $g_{j}$ to the average correlation in
the Heisenberg chain and \( C^{xx}(r_{j}) \) in the \emph{XX} chain
given by a geometric series in \( 2\gamma ^{2}_{3} \). For the Heisenberg
chain \( 2\gamma ^{2}_{3}=\frac {8}{9}<1 \), and thus \begin{equation}
\label{corrheisrs}
C(r_{j})\sim g_{j}n_{j}=\sigma \left| c_{0}\right| r^{-1}_{j},
\end{equation}
where \( c_{0} \) is the correlation between end spins in a three-spin
block and \( n_{j}=\sigma /r_{j} \) is the fraction of such blocks
in the \( j \)th generation. For the \emph{XX} chain \( 2\gamma ^{2}_{3}=1 \),
so that \( C^{xx}(r_{j}) \) carries a logarithmic correction, \begin{equation}
\label{corrxxrs}
C^{xx}(r_{j})\sim g_{j}n_{j}=\left| c_{0}\right| \left( y_{0}+y_{1}\ln r_{j}\right) r^{-1}_{j},
\end{equation}
where \( y_{0} \) and \( y_{1} \) are constants. The \( zz \) correlation
between end spins in a three-spin block is zero, so that the dominant
correlations correspond to spin pairs (connected through one of the
effective end spins and the middle spin) at distances \( r^{\prime }_{j}=4^{j-1}\pm 4^{j-2}\pm 4^{j-3}\pm \cdots \pm 1 \),
with average \( \langle r^{\prime }_{j}\rangle =4^{j-1} \),
and are given by \( g^{\prime }_{j}\sim 1/2^{j-1} \). We then have
\begin{equation}
\label{corrzzrs}
C^{zz}(r^{\prime }_{j})\sim g^{\prime }_{j}n_{j}=\sigma ^{\prime }\left| c^{\prime }_{0}\right| \left\langle r^{\prime }_{j}\right\rangle ^{-3/2}.
\end{equation}
Equations (\ref{corrxxrs}) and (\ref{corrzzrs}) should be contrasted with
the random-singlet isotropic result \( C(r)\sim r^{-2} \), indicating
a clear distinction between the ground-state phases induced by disorder
and aperiodicity, even in the presence of similar geometric fluctuations.
This is related to the inflation symmetry of the AS's,
which is absent in the random-bond case (or in aperiodic systems with
random perturbations \cite{arlego02}). Its effects are exemplified by the
fractal structure of the GS correlations visible in Fig. \ref{figrscorr},
which displays results from numerical implementations of the MDH method
for both \emph{XX} and Heisenberg chains, showing conformance to the
scaling forms in Eqs. (\ref{corrheisrs})-(\ref{corrzzrs}).

In summary, we have used the Ma-Dasgupta-Hu method to investigate
numerically and analytically the low-energy properties of aperiodic
antiferromagnetic \emph{XXZ} chains. From a general scaling argument,
we suggest that the effects of binary aperiodicity on the whole anisotropy
regime from the \emph{XX} to the Heisenberg limits can be classified
based on the same wandering exponent \( \omega  \) which is know
exactly to govern the scaling behavior of aperiodic \emph{XX} chains.
We have also shown that ground-state correlations
are dominated by characteristic distances related to the rescaling
factor of the sequences.

This work has been supported by the Brazilian agencies CAPES and FAPESP.
The author is indebted to T. A. S. Haddad for fruitful conversations.

\emph{Note added.} While this Letter was under review, some of the results for the Fibonacci chain were published in a Letter by K. Hida \cite{hida04}.

%\bibliographystyle{apsrev}
%\bibliography{/home/apvieira/latex/bibtex/mestrado.bib}

\begin{thebibliography}{17}
\bibitem[{\citenamefont{note4}()}]{note4}
\bibinfo{note}{If \(J_{a}>J_{b}\), effective spins form in the first lattice sweep, producing a Fibonacci chain with the roles of weak and strong bonds interchanged, and we are back to Fig. \ref{figfib}.}
\end{thebibliography}

\end{document}